\documentclass[sn-mathphys,iicol]{sn-jnl}

\jyear{2022}
\usepackage{url,hyperref,lineno,microtype,subcaption,xspace}
\usepackage[onehalfspacing]{setspace}

\newcommand{\TRITON}{Triton\textsuperscript{\texttrademark}\xspace}
\newcommand{\NVIDIA}{Nvidia\xspace}

\begin{document}


\title[GPUaaS in ProtoDUNE data]{Accelerating Machine Learning Inference with GPUs in ProtoDUNE Data Processing} 

\author[1]{\fnm{Tejin} \sur{Cai}}
\author*[2]{\fnm{Kenneth} \sur{Herner}}\email{kherner@fnal.gov}
\author[2]{\fnm{Tingjun} \sur{Yang}}
\author[2]{\fnm{Michael} \sur{Wang}}
\author[2]{\fnm{Maria} \sur{Acosta Flechas}}
\author[3]{\fnm{Philip} \sur{Harris}}
\author[2]{\fnm{Burt} \sur{Holzman}}
\author[2]{\fnm{Kevin} \sur{Pedro}}
\author[2]{\fnm{Nhan} \sur{Tran}}

\affil[1]{\orgdiv{Department of Physics and Astronomy}, \orgname{York University}, \orgaddress{\street{4700 Keele Street}, \city{Toronto}, \postcode{M3J 1P3}, \state{ON}, \country{Canada}}}

\affil[2]{\orgname{Fermi National Accelerator Laboratory}, \orgaddress{\street{Kirk Road and Pine Streets}, \city{Batavia}, \postcode{60510}, \state{IL}, \country{USA}}}
\affil[3]{\orgdiv{Department of Physics}, \orgname{Massachusetts Institute of Technology}, \orgaddress{\street{77 Massachusetts Avenue}, \city{Cambridge}, \postcode{02139}, \state{MA}, \country{USA}}}

\abstract{
We study the performance of a cloud-based GPU-accelerated inference server to speed up event reconstruction in neutrino data batch jobs. Using detector data from the ProtoDUNE experiment and employing the standard DUNE grid job submission tools, we attempt to reprocess the data by running several thousand concurrent grid jobs, a rate we expect to be typical of current and future neutrino physics experiments. We process most of the dataset with the GPU version of our processing algorithm and the remainder with the CPU version for timing comparisons. We find that a 100-GPU cloud-based server is able to easily meet the processing demand, and that using the GPU version of the event processing algorithm is two times faster than processing these data with the CPU version when comparing to the newest CPUs in our sample. The amount of data transferred to the inference server during the GPU runs can overwhelm even the highest-bandwidth network switches, however, unless care is taken to observe network facility limits or otherwise distribute the jobs to multiple sites. We discuss the lessons learned from this processing campaign and several avenues for future improvements.
}

\keywords{machine learning, heterogeneous (CPU+GPU) computing, GPU (graphics processing unit), particle physics, cloud computing (SaaS), neutrino physics, distributed computing}

\maketitle

\section{Introduction}

Machine learning (ML)-based algorithms have been widely used in the field of neutrino physics, for applications ranging from data acquisition to data reconstruction and analysis~\citep{Psihas:2017yuc, MINERvA:2018smv, Racah:2016gnm, Abbasi:2021ryj}. A detector technology ideally suited for computer vision applications in neutrino physics is that of liquid argon time projection chambers (LArTPCs), which are employed by the Deep Underground Neutrino Experiment (DUNE)~\citep{DUNE:2020lwj} and Short-Baseline Neutrino~\citep{MicroBooNE:2015bmn} experiments. ML applications are now deeply integrated into the event reconstruction and data analyses for the LArTPC experiments~\citep{MicroBooNE:2021bcu, ArgoNeuT:2021xtd, DUNE:2022fiy}.

The basic unit of data is a trigger record, also known as an event or event record, which consists of a series of time samples of detector readout channels at a fixed interval within a total specified time window. The number of channels, sampling rate, and readout window vary by experiment. Event record sizes for the current generation of LArTPC experiments are typically ${\leq}1$~GB and are expected to increase in the next few years.
 With increased event size, the event reconstruction, especially the inference of ML algorithms, will become a challenge. Additionally, neutrino detectors are sensitive to neutrinos from a core-collapse supernova in or near the Milky Way. One of DUNE's physics goals is to rapidly reconstruct detector trigger records from such a supernova to provide rapid localization information to optical telescopes, placing a premium on short event reconstruction times. We have demonstrated GPU-accelerated ML inference as a service, which significantly reduced the reconstruction time for simulated neutrino events in the ProtoDUNE experiment~\citep{Wang:2020fjr}. Later, we tested the same GPU-as-a-Service (GPUaaS) approach to process the entire ProtoDUNE Run I dataset to demonstrate the scalability of this method. This paper reports the results of those tests.

\section{Infrastructure setup and methods}
\subsection{ProtoDUNE description}

The ProtoDUNE single phase detector (ProtoDUNE-SP)~\citep{DUNE:2020cqd,DUNE:2021hwx} is a liquid argon time projection chamber (LArTPC) that serves as a prototype for the first far detector module of DUNE~\citep{DUNE:2020lwj}. The ProtoDUNE-SP is installed at the CERN Neutrino Platform~\citep{Pietropaolo:2017jlh}. It has an active volume of $7.2\times6.1\times7.0$ m$^{3}$. The TPC wires are read out by 15,360 electric channels at a rate of 2 MHz. A typical event record consists of 6000 time samples, corresponding to a 3\,ms time window. Between October 10 and November 11, 2018, ProtoDUNE-SP was exposed to a beam that delivers charged pions, kaons, protons, muons and electrons with momenta in the range 0.3 GeV/$c$ to 7 GeV/$c$. After the beam runs ended, ProtoDUNE-SP continued to collect cosmic ray and calibration data until July 20, 2020, after which the detector decommissioning started. The total number of trigger records during the beam period, which consist of both beam interactions and non-beam interactions such as cosmic rays, is approximately 7.2 million.

\begin{figure*}[htbp!]
    \centering
    \includegraphics[width=0.95\textwidth]{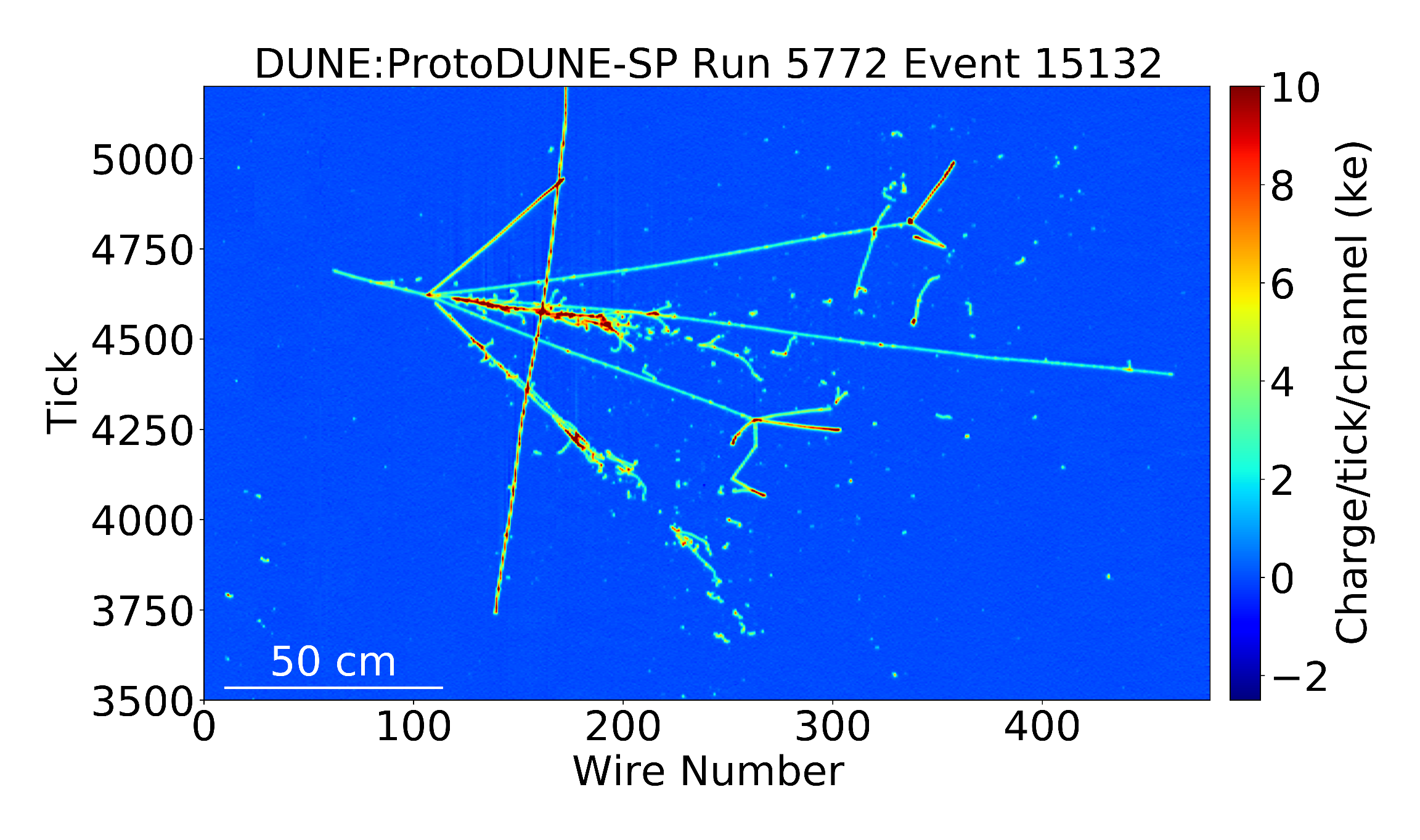}
    \caption{A 6 GeV/$c$ beam $\pi^+$ interaction in the ProtoDUNE-SP detector~\cite{DUNE:2020cqd}. The x axis shows the wire number. The y axis shows the time tick in the unit of 0.5 $\mu$s. The color scale represents the charge deposition.}
    \label{fig:r5772_e15132}
\end{figure*}

A ProtoDUNE-SP TPC waveform recorded by a single electric channel consists of both signals and noise. There are typically three sources of signals. During the beam runs, the beam particles can interact with the liquid argon inside the TPC and produce both ionization electrons and scintillation light. Since ProtoDUNE-SP is located on the Earth’s surface, it is subject to a large flux of cosmic ray muons, which induce signals over the entire detector. There are also radioactive backgrounds such as $^{39}$Ar that generate low energy signals on the scale of a few hundred keV to a few MeV. Figure~\ref{fig:r5772_e15132} shows the event display of a 6 GeV/$c$ pion interaction in the ProtoDUNE-SP detector.

The first step in the reconstruction of events in the TPC is the signal processing.  The goal of this stage is to produce distributions of charge arrival times and positions given the input TPC waveforms. The effects of induced currents
due to drifting and collecting charge, as well as the response of the front-end electronics, are removed through de-convolution. The charge arrival distributions are
used in subsequent reconstruction steps, starting with hit finding. The hit finding algorithm fits peaks in the wire waveforms, where a hit represents a charge deposition on a single wire at a given time. Each hit corresponds to a fitted peak. The hits are input to pattern recognition algorithms such as Pandora~\citep{pandorasdk,pandorauboone,DUNE:2022wlc}. This stage finds the high-level objects associated with particles, like tracks, showers, and vertices, and assembles them into a hierarchy of parent-daughter nodes that ultimately point back to the candidate neutrino interaction. More details on the reconstruction workflow are described in Ref.~\cite{DUNE:2020cqd}.

In ProtoDUNE-SP, a novel algorithm is developed based on a convolutional neural network (CNN) to perform the classification of each reconstructed hit as track-like or arising from electromagnetic cascades~\citep{DUNE:2022fiy}. These hit-level classifications can be used alongside pattern recognition based reconstruction algorithms such as Pandora to refine the track or shower classification of reconstructed particles. The CNN model was trained using TensorFlow~\citep{tensorflow2015-whitepaper}. In the DUNE code base this algorithm is known as {\it EmTrkMichelId}; hereafter, we call this algorithm EmTrk.

In order to improve the efficiency and speed of the inference of ML algorithms in a large-scale data processing, GPU acceleration specifically for the ProtoDUNE-SP reconstruction chain has been integrated without disrupting the native computing workflow using the services for optimized network inference on coprocessors (SONIC) approach~\citep{larrecodnn,Wang:2020fjr}. With the integrated framework, the most time-consuming task, track and particle shower hit identification, runs faster by a factor of 17. This results in a factor of 2.7 reduction in the total processing time when compared with CPU-only production. This initial test using a small number of simulated ProtoDUNE-SP events showed a viable, cost-effective way to solve the computing challenge facing the neutrino experiments. In this work, we report the results of reprocessing the entire 7 million ProtoDUNE-SP events taken during the test beam runs with the SONIC-enabled framework.

\subsection{Inference server setup}

The \NVIDIA \TRITON Inference Server is an open-source inference serving software that helps standardize model deployment and execution; its goal is to deliver fast and scalable AI in production~\citep{NVIDIA_triton_rt}. NVIDIA provides multiple ways to deploy the inference server on different cloud providers and infrastructure types, including both bare metal and containerized workloads.

\begin{figure*}[htp]
\centering\centering
\includegraphics[width=0.6\textwidth]{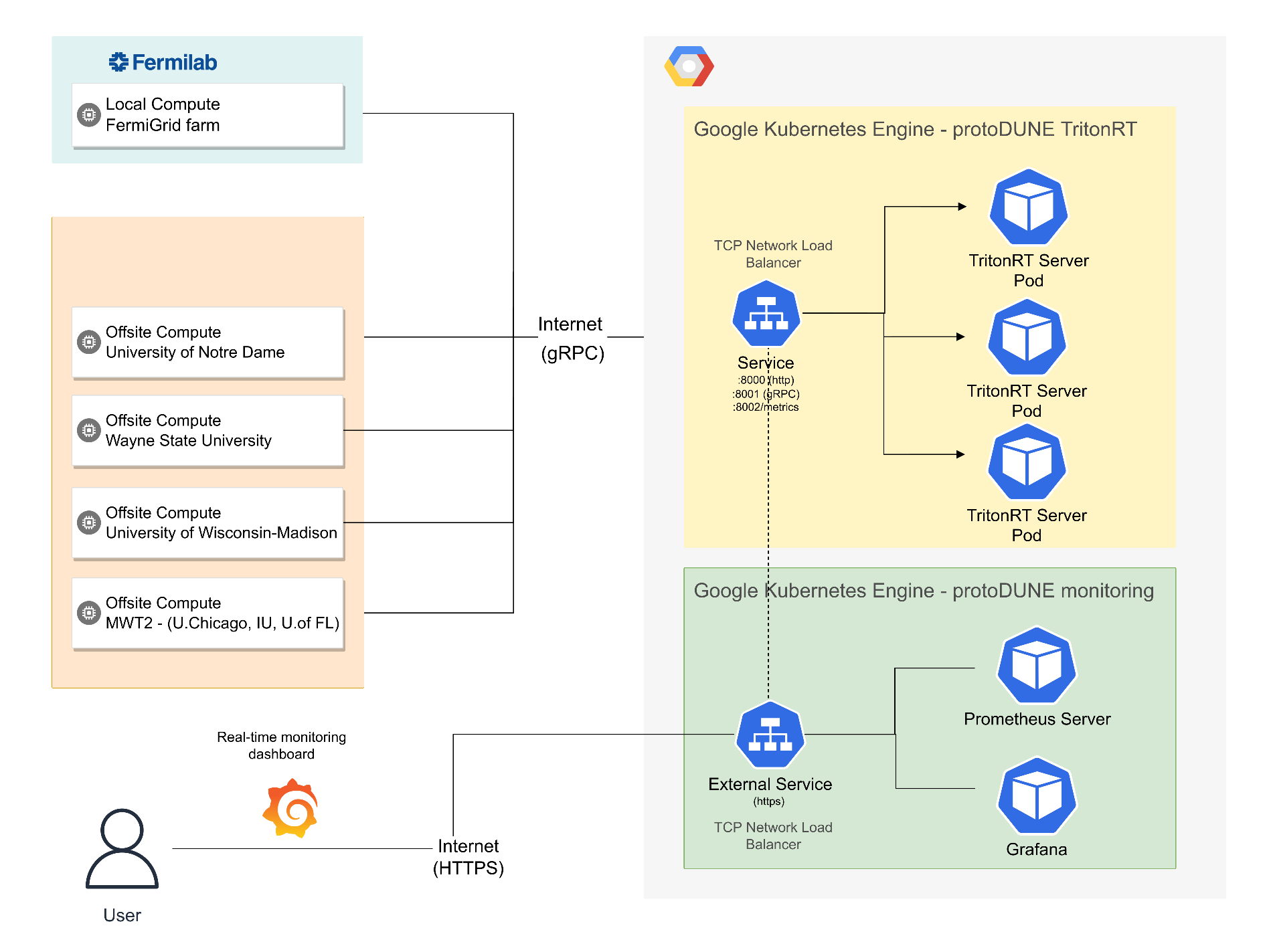}
\caption{ProtoDUNE GPUaaS component diagram depicting local and remote batch inference runs submitted from Fermilab and OSG Grid sites.}
\label{fig:ProtoDUNEGPUaaScloud}
\end{figure*}

This study uses a cloud-based deployment of \NVIDIA \TRITON Inference Server within a Google Cloud Kubernetes Engine~\citep{Google_KubernetesE} cluster on virtual infrastructure provided by Google Cloud Platform. The use of this technology enables us to deploy a flexible GPUaaS model where a public endpoint takes remote inference requests from various geographically distributed sources as depicted in Figure~\ref{fig:ProtoDUNEGPUaaScloud}. The Triton\textsuperscript{\texttrademark} server running on the Google cloud supports different backends. We use the TensorFlow (version 1.15.5) backend for the inference of the EmTrk algorithm.

In a similar way as Ref.~\citep{Wang:2020fjr}, this study uses several \TRITON servers split into separate Kubernetes deployments with common services for networking and external load balancing in the form of ingress objects~\citep{Google_KubernetesE_ingress}. One significant improvement for the current study is the deployment of metrics and monitoring which provided us with observability within the system in different states. In IT and cloud computing, observability is the ability to measure a system's current state based on the data it generates, such as logs, metrics, and traces. It relies on telemetry derived from instrumentation that comes from the endpoints and services in computing environments. \TRITON provides a built-in metrics endpoint~\citep{NVIDIA_triton_rt_metrics} that publishes plain-text data in Prometheus format~\citep{Prometheus_formats}.

\subsection{Methods}

The DUNE collaboration undertook a production campaign in 2021 to process ProtoDUNE-SP data using the LArSoft toolkit~\citep{Snider:2017wjd} version v09\_30\_00. Each production run during the beam period comprises several data files, each containing between 100 and 150 data records. In contrast to the previous work, in which DUNE simulation events were processed by submitting jobs locally to a dedicated queue, we submit jobs to process each file via the current standard DUNE workflow management and job submission systems~\citep{duneprod,POMSCHEP}, thus requiring no special treatment. Jobs may run either at Fermilab or one of several remote sites that we reach with opportunistic access enabled by the OSG Consortium~\citep{OSG2007}.

We begin from the existing reconstructed outputs and apply the updated EmTrk algorithm to produce new outputs. Of the 7.2 million ProtoDUNE events during the 2018 beam period, we process 6.4 million through the SONIC infrastructure, and 800k with the CPU-only version of the same algorithm for comparison. The OSG sites included in the SONIC runs were chosen to be geographically proximate to the location of the Google Cloud GPU servers (which were in Iowa, USA at the time) in order to minimize the latency in data transmissions. Latency between the sites and Google Cloud server as measured by the ping utility was typically between 15 and 20 ms.

The difference in the time spent in the inference step is the primary metric with which we assess the advantage of GPUaaS over traditional CPU processing. Each job produces a log file that statistically summarizes the time spent on each stage of the event reconstruction for the job as a whole. The log has no record of per-stage processing time at the individual event level, but we can closely approximate it by taking the difference between the start times of consecutive events. We estimate the per-event EmTrk duration by subtracting the median non-EmTrk duration from the total event duration, as the non-EmTrk stages display very little time variation across events. The CNN-based hit classification occurs in the EmTrk stage and is the most time-consuming step in the event reconstruction, typically accounting for more than 90\% of the processing time.  

\section{Results}

\subsection{CPU-only runs}
We process a set of 13 runs using CPU-based TensorFlow both at Fermilab and several off-site locations. The off-site locations are the University of Notre Dame, the University of Victoria, and the high performance computing center at Wayne State University. The TensorFlow version used in the CPU-only runs is 2.3.1.  Although the TensorFlow version differs from that used in the backend for the GPU runs, the main differences between the two versions concern additional support for advanced CPU instruction sets. We therefore do not expect any significant performance differences between the two versions in the GPU case. Table~\ref{tab:cpu_offsite} summarizes the number of events processed at each site and the median processing times. We did not request any specific CPU type when submitting these jobs since typical DUNE practice is to use any and all available CPU types.

\begin{table}[h]
\renewcommand{\arraystretch}{1.3}
    \caption{List of CPU-only run sites and median processing time}
    \centering
    \begin{tabular}{crr}
       OSG Site  & N samples & Median processing time (s) \\\hline
       FermiGrid & 746603 & 79 \\
       Notre Dame  & 36082 & 68 \\
       Victoria    & 10944 & 52 \\
       Wayne State  & 4242 & 45 \\
    \end{tabular}
    \label{tab:cpu_offsite}
\end{table}

\begin{figure}
    \centering
    \includegraphics[width=.5\textwidth]{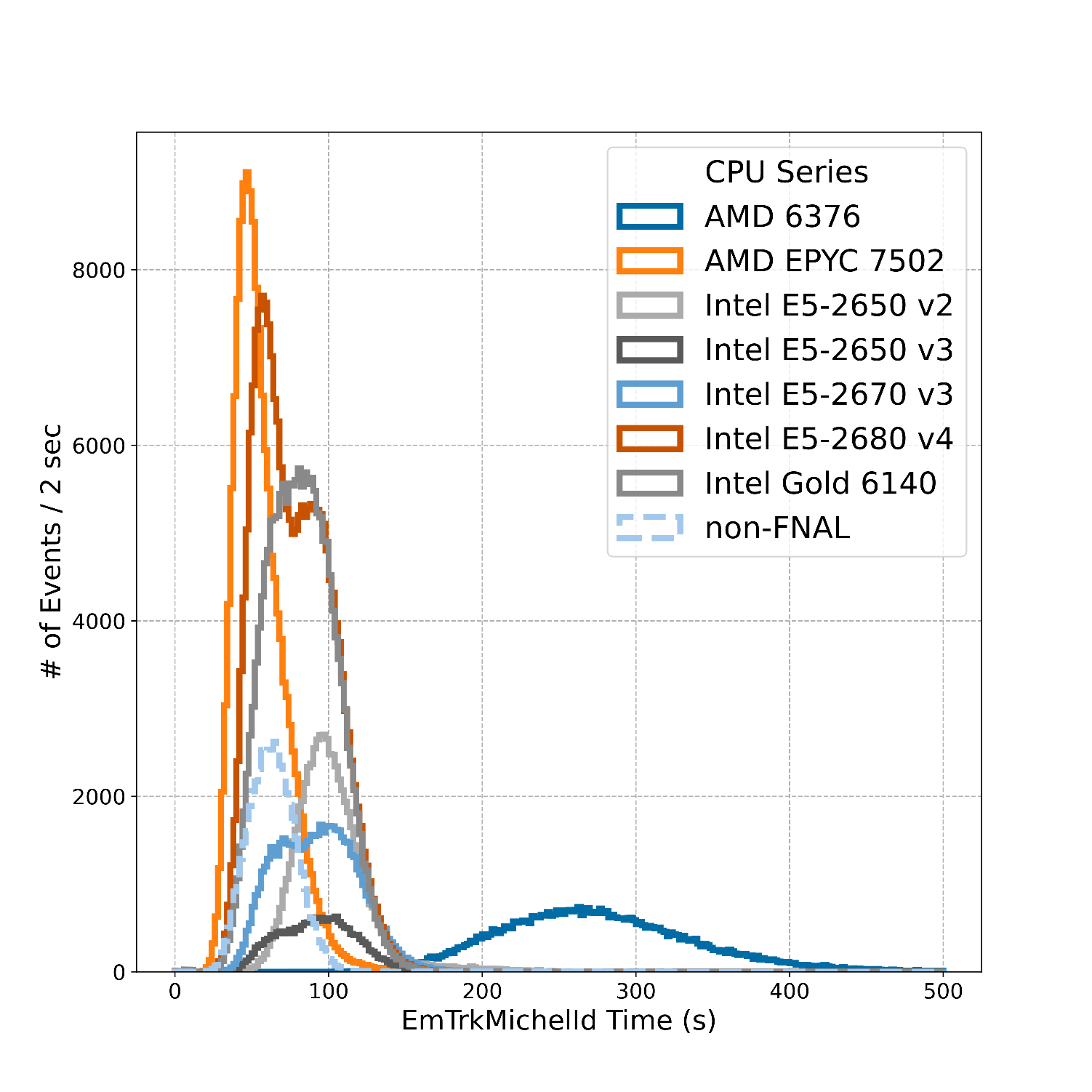}
    \caption{Timing distributions for CPU-only runs, broken down by CPU type.}
    \label{fig:cpu_series_timing}
\end{figure}

There is a clear dependence on processor type in the EmTrk processing time distribution. In general, more recent CPUs process events faster. Figure~\ref{fig:cpu_series_timing} shows the CPU-based EmTrk timing for each of the CPU types currently available on the Fermilab general purpose batch farm. We do not have access to CPU type information outside of Fermilab and thus group them together.

\subsection{GPU runs}
\begin{figure}[htp!]
\centering
\includegraphics[width=0.5\textwidth]{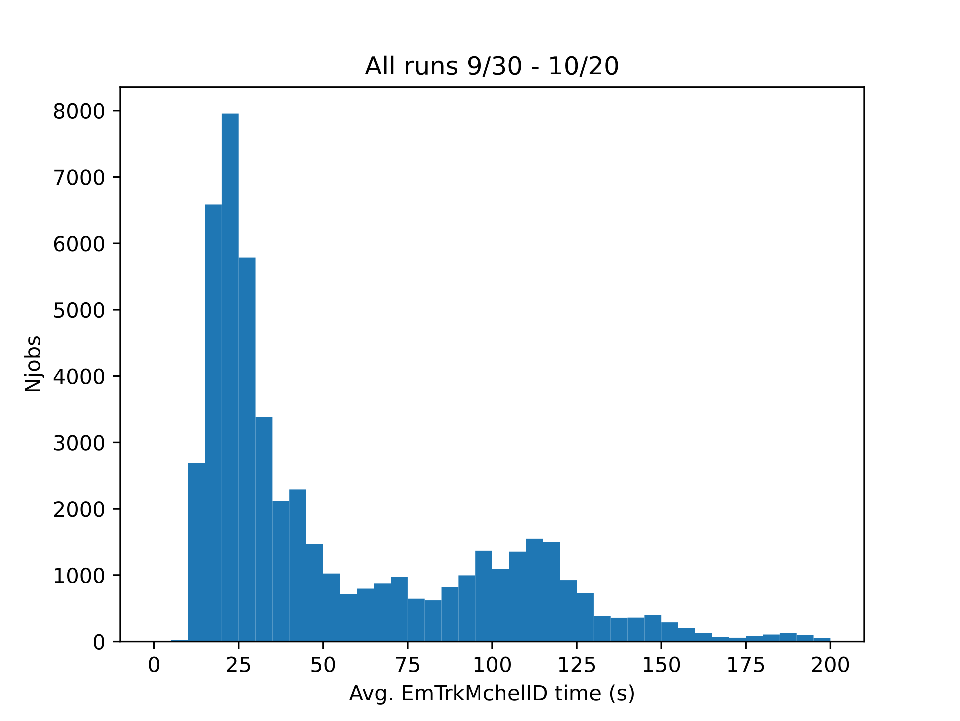}
\caption{Average EmTrk times for GPU runs during the period September 30, 2021 to October 20, 2021. The double peak structure arises from periods during which the outbound network connection from the Fermilab grid processing center was saturated.}\label{fig:GPUall}
\end{figure}

Our main processing effort uses the GPUaaS infrastructure as described. Figure~\ref{fig:GPUall} shows the average EmTrk processing time when using the GPUaaS infrastructure for our entire running period. The first peak at approximately 20 s represents a factor of two improvement with respect to the fastest CPU-only runs, and a factor of roughly 11 over the slowest CPU runs. It is important to note that the EmTrk times we report here are wall times measured within the job, and thus include contributions from network latency to and from the server. There is another peak in the distribution with a median of over 100 s, to which we now turn. 

\subsubsection{Outbound network saturation}
\begin{figure}[htp!]
\centering
\includegraphics[width=0.5\textwidth]{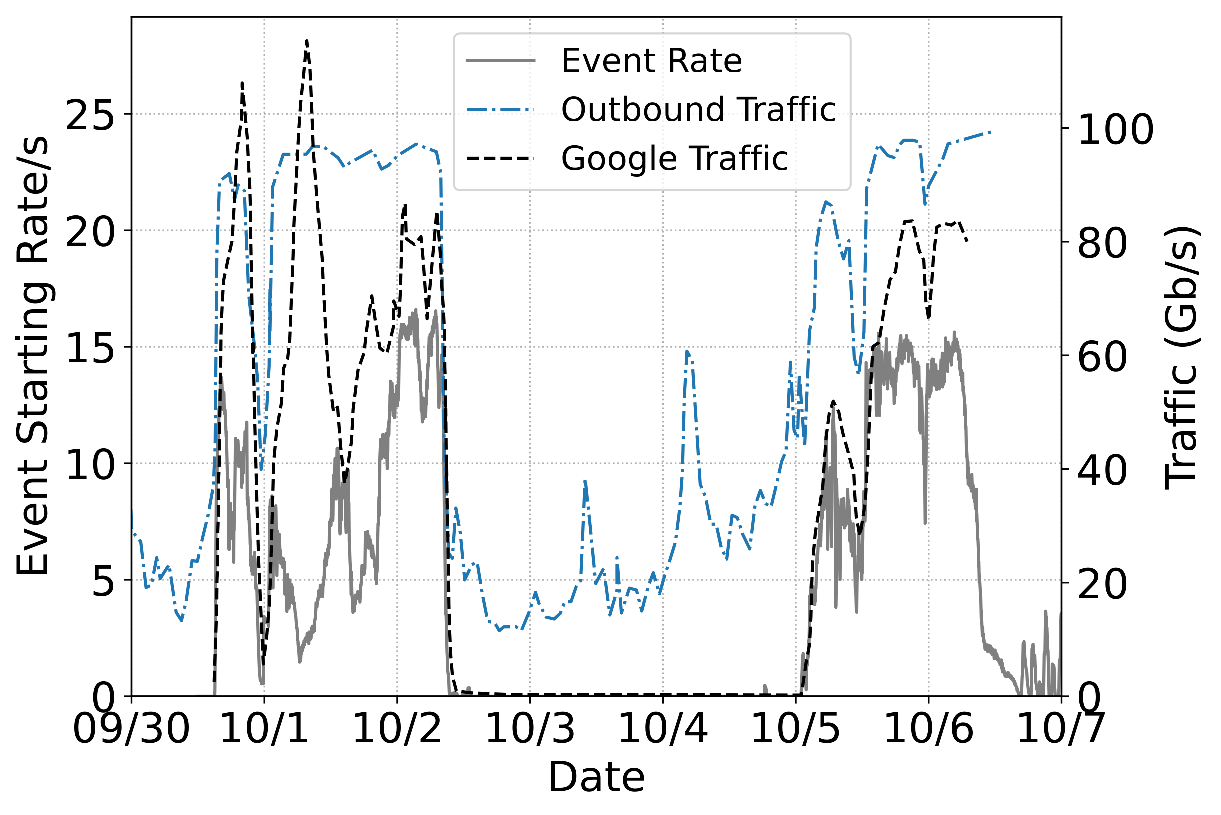}
\caption{Overlay of network traffic and event processing start rate at FermiGrid as a function of time, which is a proxy for the number of concurrent jobs. The origin day is September 30, 2021. The solid line is the event start rate, the blue dot-dash line is the outbound network traffic rate through the 100 Gb/s switch at Fermilab used by the batch processing cluster, and the black dashed line is the ingress rate to the Google cloud server.
We are unable to disambiguate traffic sources through the switch, so the blue dot-dash line represents the total traffic as opposed to only traffic generated by our processing campaign. We see that the network switch was effectively saturated in multiple instances, though Google ingress was not.}\label{fig:nevt_traffic_1D}
\end{figure}

\begin{figure}[htp!]
\centering
\includegraphics[width=0.45\textwidth]{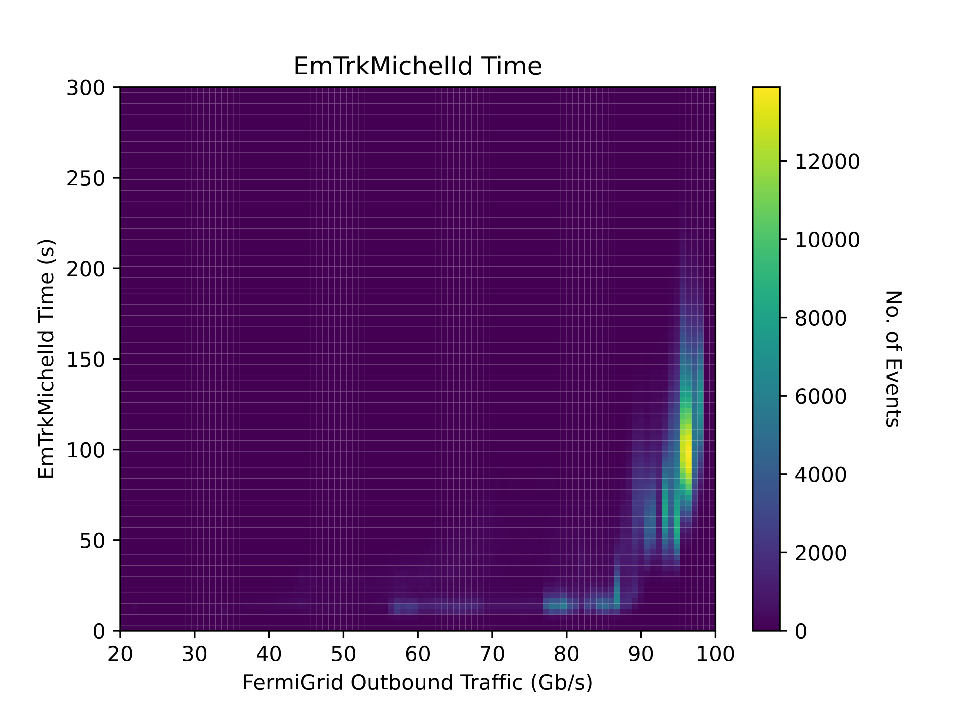}
\includegraphics[width=0.45\textwidth]{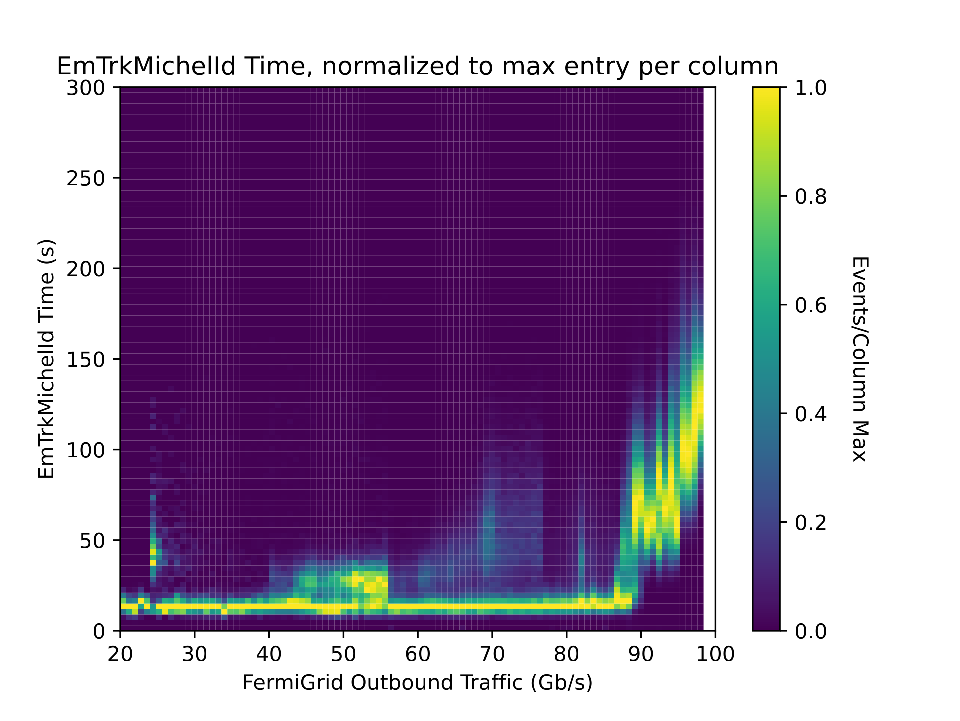}
\caption{The average EmTrk duration before Oct. 7 as a function of the total network traffic through the 100 Gb/s network switch at Fermilab used by the batch processing cluster. The top plot shows the real event rate. The bottom plot is the same as the top one, with each column scaled separately so the maximum amplitude is 1 for each column.}\label{fig:michel_traffic}
\end{figure}

\begin{figure}[htp!]
\centering
\includegraphics[width=0.5\textwidth]{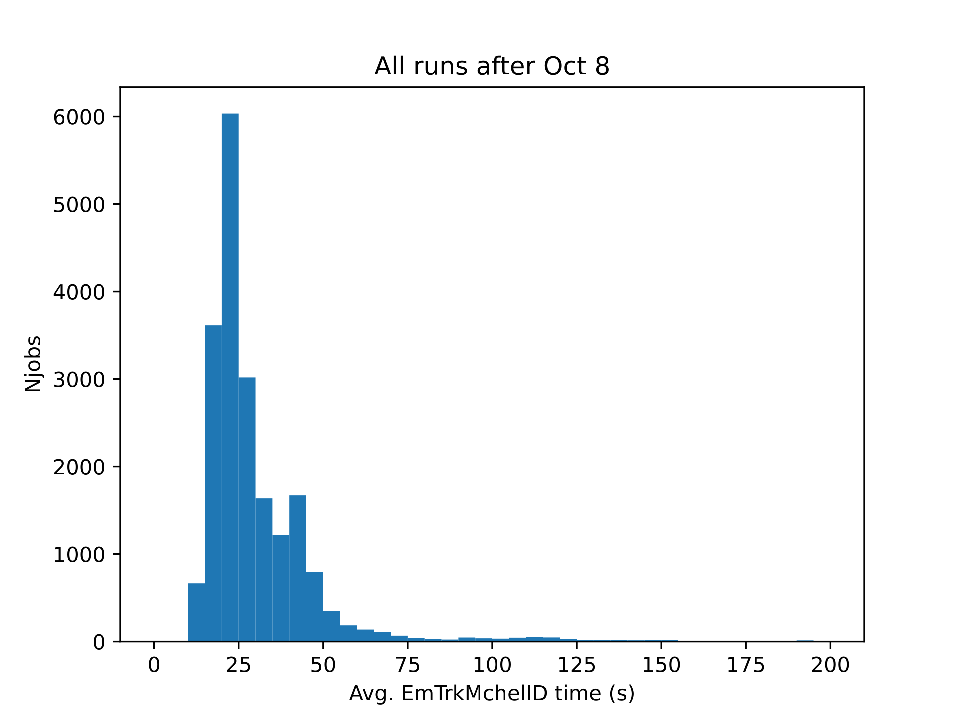}
\caption{The average time spent in the EmTrk task for all GPU jobs after October 8, when the network saturation had subsided.}\label{fig:GPUafterOct8}
\end{figure}

During the first period of GPU running we averaged between 200 and 2000 concurrent jobs. Figure~\ref{fig:nevt_traffic_1D} shows the overlay of network traffic and event processing start rate during the period of September 30, 2021 to October 6, 2021.
As the event start rate increases because of the rise in the number of concurrent jobs, we see that the 100 Gb/s outbound network connection used by the Fermilab data center where the jobs run becomes saturated. While our jobs were not solely responsible for the saturation (the connection serves the entire cluster), the saturation did result in a significant increase in the average EmTrk processing time as shown in Figure~\ref{fig:michel_traffic}. The highest job concurrency levels were on October 5, when unusually low demand for computing resources from other Fermilab experiments resulted in a large number of opportunistic job slots being available at Fermilab. We were, without any direct intervention, thus able to scale up to approximately 6,000 concurrent jobs. The monitoring does show switch saturation as early as October 1, however. After learning of the network saturation we implemented a concurrency limit on jobs of approximately 600; thereafter the jobs ran without incident and the EmTrk times returned to pre-saturation levels (see Figure~\ref{fig:GPUafterOct8}).

\section{Discussion}

\begin{figure*}[!htp]
    \centering
    \includegraphics[width=1\textwidth]{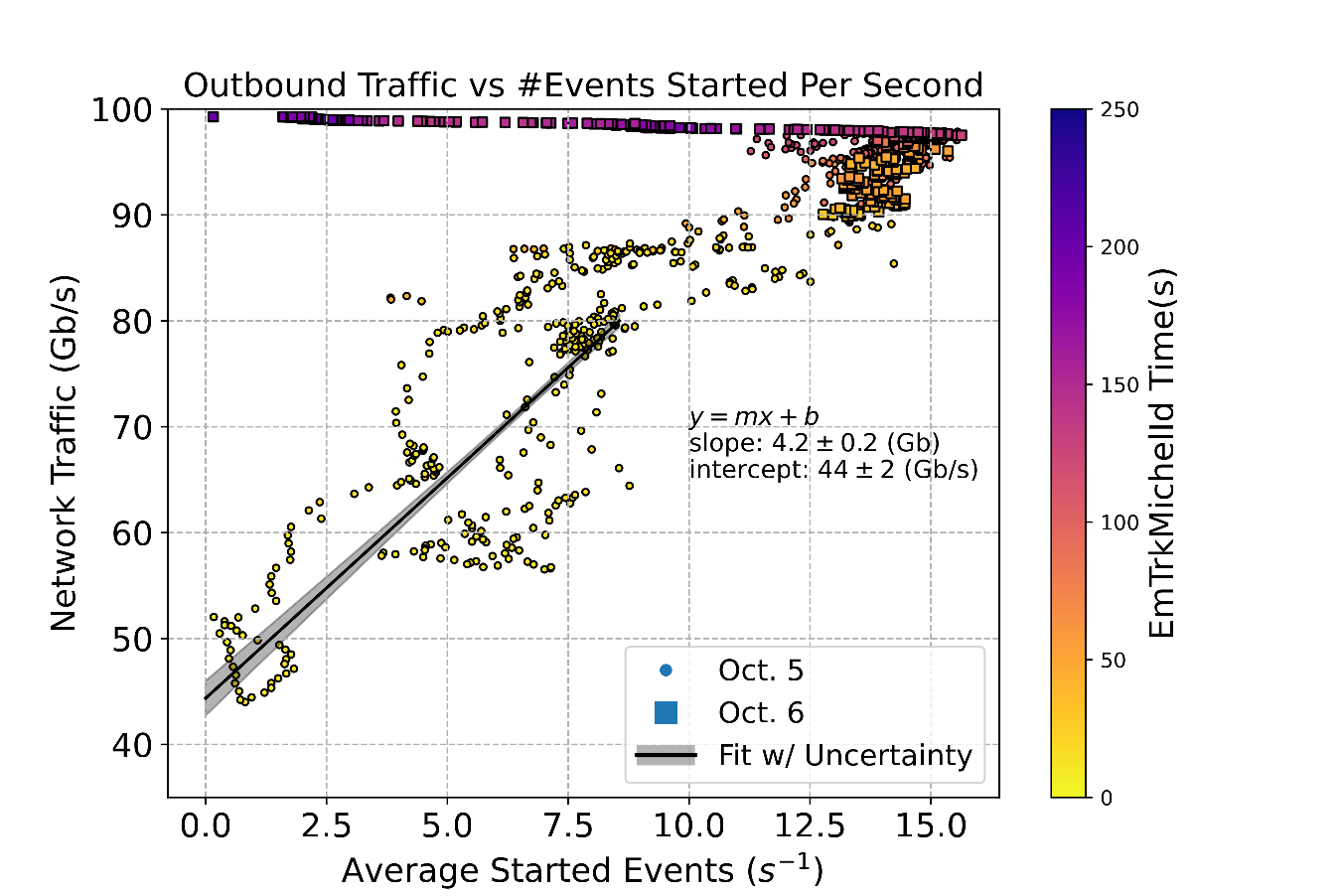}
    \caption{The outbound network traffic vs. the average event start rate per second in 2-minute sliding windows, on October 5 and October 6. Data from each day is denoted with a different marker type. The color coding corresponds to the median EmTrk time for events in each sliding window. The linear fit to the traffic below $80$ Gb/s indicates that each event sends $4.2\pm 0.2$ Gb of outbound traffic, on top of $44\pm2$ Gb/s of baseline traffic from non-ProtoDUNE sources. }\label{fig:ratevstraffic}
\end{figure*}

In order to understand the impact of ProtoDUNE jobs on the Fermilab network traffic, we plot the distribution of event processing start rate versus network traffic in Figure~\ref{fig:ratevstraffic}. Even though the network traffic has contributions from all grid jobs at Fermilab, there is a clear correlation between the number of ProtoDUNE concurrent jobs and the increase of network traffic. We fit a straight line to the data points below the network traffic of 80 Gb/s. The slope of the best fit line is $4.2\pm0.2$ Gb, which is the average outbound data transmission per event. The intercept is $44\pm2$ Gb/s, which is the average traffic from non-ProtoDUNE grid jobs. Based on the discussion of transmission time in Ref.~\citep{Wang:2020fjr}, for 55,000 inferences per event, with each input a $48 \times 48$ image at 32 bits, the total amount of data transmitted is about 4.1 Gigabits per event. This is consistent with the slope of the best fit straight line. The spread in data with respect to the straight line could be caused by the variation in the number of non-ProtoDUNE grid jobs during this period. 

Figure~\ref{fig:GPUafterOct8} indicates that the average processing time is roughly 25 s/event for the GPU jobs. Assuming the entire 100 Gb/s bandwidth is available to the ProtoDUNE jobs, the maximum number of concurrent ProtoDUNE jobs we can run without saturating the network is $(100~\text{Gb/s})/(4.1~\text{Gb/event})\cdot(25~\text{s/event}) \simeq 600$. This is consistent with the concurrency limit of 600 jobs that we implemented after October 7.

Based on the above discussions, we conclude that, while overall computational time clearly decreases using GPUaaS, one does have to take particular care to 
understand what the expected data movement requirements will be for jobs using this architecture, and to set job concurrency limits appropriate to the capabilities of each local computing site and input data source. HTCondor~\cite{condorgrid,BOCKELMAN2020101213} in particular has the ability to define an arbitrary kind of resource that each job requires; one could define a ``bandwidth'' resource for these jobs, for example. HTCondor additionally allows configuring the job submissions to prevent more jobs to start at a given site once the sum of consumed resources by running jobs at that site reaches a certain threshold. Therefore, if one knows the total network capacity of each site hosting jobs, one can configure per-site job limits and prevent network saturation in an automated way. 

\subsection{Future improvements}
A number of improvements to overall scalability and ease of use are possible. In addition to automatic job concurrency limits to prevent network saturation as previously described, we are exploring the possibility of compressing the data sent to the GPU server to reduce the overall bandwidth requirements. While a reduced payload would obviously increase job concurrency limits, that must be balanced against the additional run time that would be introduced in compressing and decompressing the data on the worker node and server, respectively. Another desirable area of improvement is in overall ease of use and human effort requirements. In the current setup we make use of the standard DUNE Production job submission infrastructure, which allows for a high degree of automated job submission, but due to the current nature of the cloud server it requires an authorized individual to manually instantiate the GPU inference server before we submit jobs. Establishing a method of automatically instantiating the server at job submission time and automatically ramping it down when the associated jobs are complete would avoid a clear possible failure point should no authorized individuals be available when the infrastructure is needed.

A second option to study is to use several geographically distributed inference servers instead of a single server, while also spreading the job workload over a much broader range of sites. Expanding the site pool has the advantage of making it much less likely that any single site would get enough work assigned to saturate its external connectivity, and using several inference servers spread around the world would help to mitigate the potential problem of network latency becoming comparable to the inference time. The cost changes in this scenario (for example, the relative cost of three cloud servers versus a single server three times the size) must be assessed and taken into account. Another consideration is how the overall event processing times would change if the worker nodes were much more geographically diffuse than they were for this study. Since we stream the input data over the network, longer network paths between the worker nodes and input data sources may lead to the non-EmTrk portions of the event processing taking longer, which in turn affects the total event processing time. DUNE is able to distribute data to various storage elements around the world via the Rucio framework~\citep{Barisits2019}, and pre-placing the data of interest at storage elements close to the sites to be used for processing may mitigate such concerns, though it is not required.

Another potential avenue is to use the GPU server infrastructure, but to use sites with GPUs available on the worker nodes, and run an independent server on each worker node.
Several high-performance computing sites have built or are building clusters with readily available GPUs, and in some cases with multiple GPUs on each worker node, that would naturally lend themselves to such a setup. 
If the jobs run on worker nodes with local GPUs, external network connectivity limitations become unimportant for carrying out the inference calculations.
In fact, \TRITON allows the use of shared memory for direct data transfer between CPU and GPU when the GPU is local. While it may not be necessary to retain the server infrastructure in these cases, the advantage of doing so is that the experiment software does not have to be modified to directly access the GPU, making it maximally portable and easier to maintain. We plan to conduct a similar study using this type of setup in the future. 

\section{Summary}

We have reprocessed approximately seven million data events from the ProtoDUNE detector installed at CERN. We use an \NVIDIA \TRITON inference server hosted on the Google Cloud Platform to run the most computationally expensive step of the workflow on a GPU, speeding up the required processing time by more than a factor of two, even comparing to the fastest CPU runs. Running at a scale similar to that expected during regular ProtoDUNE-II and DUNE operations, we see the expected performance improvement until the network switch through which the majority of our jobs communicate becomes saturated. Despite that, the cloud infrastructure easily kept up with demand and demonstrates the viability of the GPUaaS model at a level sufficient for current and future high-energy physics experiments, as long as the job concurrency levels at each site respect the site's network resource limits. With several promising avenues of improvement to explore, we expect that this computing model will become even more capable and easier to use in the future.

\section*{Author Contributions}
All authors contributed to the study conception and design. Material preparation, data collection and analysis were performed by Tejin Cai, Kenneth Herner, and Tingjun Yang. The first draft of the manuscript was prepared by Tejin Cai, Maria Acosta Flechas, Kenneth Herner, Kevin Pedro, Nhan Tran, and Tingjun Yang. All authors read and approved the final manuscript.

\section*{Acknowledgments}

We acknowledge the Fast Machine Learning collective as an open community of multi-domain experts and collaborators. This community was important for the development of this project. We acknowledge the DUNE collaboration for providing the ProtoDUNE-SP code base and data samples. The analysis is enabled in part by the Digital Research Alliance of Canada.

\section*{Declarations}
\subsection*{Competing Interests}
The authors have no competing interests to declare that are relevant to the content of this article.
\subsection*{Data Availability}
The datasets generated during and/or analysed during the current study are available from the corresponding author on reasonable request.
\subsection*{Funding}
MF, KH, BH, KP, NT, MW, and TY are supported by Fermi Research Alliance, LLC under Contract No. DE-AC02-07CH11359 with the U.S. Department of Energy, Office of Science, Office of High Energy Physics. NT is partially supported by the U.S. Department of Energy Early Career Award. KP is partially supported by the High Velocity Artificial Intelligence grant as part of the U.S. Department of Energy High Energy Physics Computational HEP program. PH is supported by NSF grants \#1934700, \#193146.  Cloud credits for this study were provided by Internet2 managed Exploring Cloud to accelerate Science (NSF grant PHY-190444). TC is supported by NSERC Canada. 

\bibliography{test}

\end{document}